\documentclass[lettersize,journal]{IEEEtran}
\usepackage{amsmath,amsfonts}
\usepackage{array}
\usepackage[caption=false,font=footnotesize,labelfont=sf,textfont=sf]{subfig}
\usepackage{textcomp}
\usepackage{stfloats}
\usepackage{url}
\usepackage{verbatim}
\usepackage{graphicx}
\usepackage{multirow}
\usepackage{cite}
\usepackage{xcolor}
\usepackage{algpseudocode}
\usepackage[linesnumbered,ruled]{algorithm2e}
\usepackage{makecell}

\begin{document}

\title{FPGAs (Can Get Some) SATisfaction}

\author{ \IEEEauthorblockN{H. Govindasamy\IEEEauthorrefmark{1}, B. Esfandiari\IEEEauthorrefmark{1}, P. Garcia\IEEEauthorrefmark{2}}
    \\\IEEEauthorblockA{\IEEEauthorrefmark{1}\{hari, babak\}@sce.carleton.ca, Dpt. Systems Computer Engineering, Carleton University, Canada}
    \\\IEEEauthorblockA{\IEEEauthorrefmark{2}paulo.g@chula.ac.th, International School of Engineering, Chulalongkorn University, Thailand
    }}

\markboth{IEEE Embedded Systems Letters}%
{Govindasamy \MakeLowercase{\textit{et al.}}: FPGAs (Can Get Some) SATisfaction}

\IEEEpubid{0000--0000/00\$00.00~\copyright~2021 IEEE}

\maketitle

\begin{abstract}
 We present a hardware-accelerated SAT solver suitable for processor/Field Programmable Gate Arrays (FPGA) hybrid platforms, which have become the norm in the embedded domain. Our solution addresses a known bottleneck in SAT solving acceleration: unlike prior state-of-the-art solutions that have addressed the same bottleneck by limiting the amount of exploited parallelism, our solver takes advantage of fine-grained parallelization opportunities by hot-swapping FPGA clause assignments at runtime. It is also the first modern completely open-source SAT accelerator, and formula size is limited only by the amount of available external memory, not by on-chip FPGA memory.
\par Evaluation is performed on a Xilinx Zynq platform: experiments support that hardware acceleration results in shorter execution time across varying formula sizes, subject to formula partitioning strategy. We outperform prior state-of-the-art by 1.7x and 1.1x, respectively, for 2 representative benchmarks,  and boast up to 6x performance increase over software-only implementation.
\end{abstract}


\begin{figure*}[b!]
\includegraphics[width=0.9\textwidth]{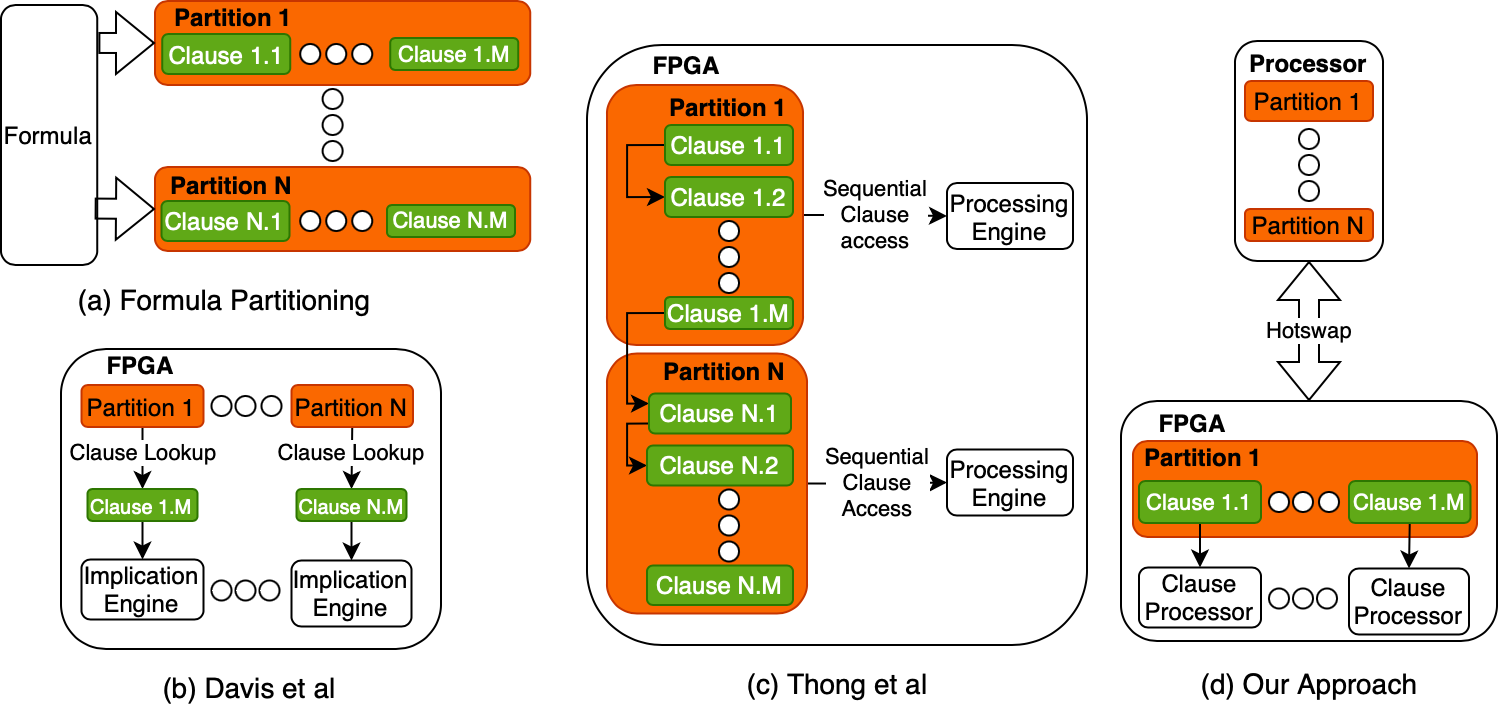}
\centering
\caption{(a) Formula partitioning. (b) Formula stored directly on FPGA; single clause retrieved for processing per partition. (c) Formula stored directly on FPGA; partitions connected in partial order. Clauses processed sequentially. (d) Formula stored in external memory ("software" view) and partitions hot-swapped as required. Clauses are mapped directly to clause engines.} 
\label{fig:arch_comparison}
\end{figure*}

\section{Introduction}
\IEEEPARstart{E}{fficient} computation of Boolean Satisfiability problem (SAT) is extremely important across several engineering domains, as SAT can encode all NP-complete problems \cite{10.5555/574848, cook2023complexity}. In the power/performance constrained embedded world, SAT solvers were not feasible \cite{zhong1998accelerating} until the recent advent of sufficiently powerful Field Programmable Gate Arrays (FPGA) offering acceleration opportunities \cite{zhong1999using}.
\par The state of the art on SAT solver acceleration is comprised of either incomplete solvers \cite{8697729} or complete solvers with only coarse-grained parallelization \cite{6691124, 10.1145/1497561.1497576}: we point interested readers to a comprehensive survey of hardware SAT solvers \cite{SOHANGHPURWALA2017170}. Our work directly extends (and is compared with) works by Davis et al \cite{4555925} and Thong et al \cite{6691124}, both hardware-accelerations of Boolean Constraint Propagation (BCP), a sub-component of the Davis-Putnam-Logemann-Loveland (DPLL) SAT solver algorithm.
\par Figure \ref{fig:arch_comparison} summarizes Davis et al, Thong et al and our approach. Davis et al's \cite{4555925} approach groups \emph{clauses} (see Section \ref{sec:background} below) across parallel accelerators, such that no two clauses on the same accelerator share variables. Using a two stage process, the variable assignment's corresponding clause is first retrieved (we point readers to \cite{4555925} for an in-depth explanation of their look-up technique), and the resulting implication (constraints on variable values in function of prior clauses) is then calculated.
\par Thong et al \cite{6691124} identify that clause look-up in Davis et al's \cite{4555925} work is slower than direct access; thus, they evaluate a clause encoding scheme that links clauses with shared variables on BRAM. The result is a sequential BCP processor with multi-threaded software support and a more performant solution when compared to Davis et al. BCP processors are replicated in hardware to maximize FPGA usage.
\par In contrast with Thong et al, we eliminate the need for clause look-up in Davis et al's \cite{4555925} work by allowing clauses to share variables within the same partition, keeping formula in external memory (i.e., in software) and hot-swapping hardware assignment of clauses at runtime, as required for implication calculation. This allows us to maintain the same high degree of parallelization as Davis et al, unlike Thong et al's sequential solution, at the expense of more swapping overhead. Experiments suggest swapping overhead can be negligible, compared to acceleration gains, subject to formula partitioning: results show our solution can outperform related work if partitioning results in little swapping, and formula size is limited only by the size of on-board memory (external RAM, not on-chip FPGA memory). Our solution is available here\footnote{\url{https://github.com/harigovind1998/FPGA_BCP_acceleration}}, deployed on the Zynq chip, the processor-FPGA platform that has become the \textit{de facto} solution across the embedded domain \cite{manev2019unexpected}; to the best of the authors' knowledge, this is the first modern open-source SAT solver accelerator.

\begin{figure*}[t!]
\includegraphics[width=0.95\textwidth]{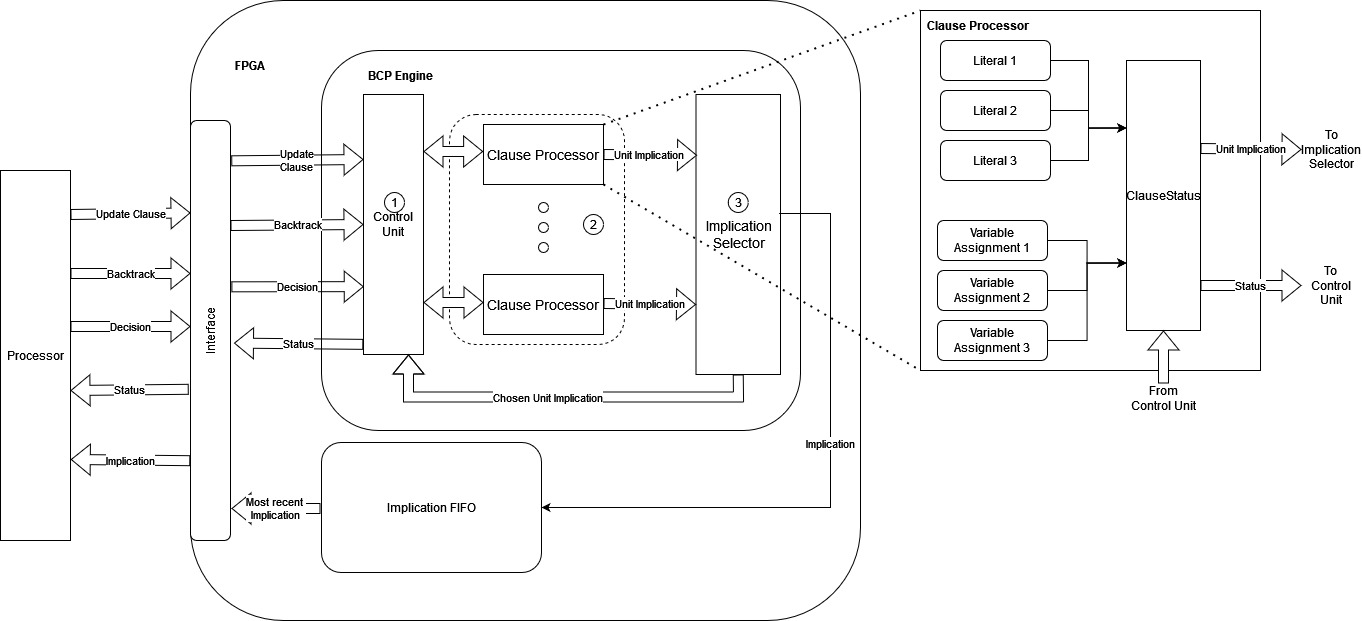}
\centering
\caption{SAT Accelerator Processor-FPGA Interface. BCP Engine accelerates part of the software SAT solving algorithm through fine-grained parallelization across Clause Processors.}
\label{fig:processor_fpga_interface}
\end{figure*}

\section{The SAT solver Architecture}

\subsection{Background: DPLL and BCP}\label{sec:background}

Given a Boolean formula, a SAT solver determines if a satisfying assignment exists; that is, does a free variable assignment exist, such that the formula evaluates to true? Boolean formulas are expressed in Conjunctive Normal Form (CNF), where the input is a conjunction of clauses, each a disjunction of \textit{literals} (an occurrence of a variable or its negation).
\par Incomplete solvers follow a greedy approach, often employing variations of Stochastic Local Search (SLS) \cite{SOHANGHPURWALA2017170}. These solvers are typically quicker than complete ones; however, satisfying assignments can be eluded and results are not guaranteed. Complete solvers, on the other hand, search exhaustively, evaluating every possible assignment combination until a satisfying assignment is found. A formula is unsatisfiable if no satisfying assignment is found after checking all combinations.
\par Many complete solvers are derivations of the DPLL algorithm \cite{SOHANGHPURWALA2017170}, proposing improvements to its decision heuristic and branch pruning techniques. DPLL consists of two stages: decision and Boolean Constraint Propagation (BCP). During decision, variables are assigned truth values, which are then propagated and evaluated during BCP, until a formula is deemed satisfiable or not. The BCP component is substantially expensive, taking up to 80-90\% of DPLL's total execution time \cite{6691124}. Both most relevant prior works (\cite{4555925},\cite{6691124}) as well as our own, are based on accelerating BCP.

\subsection{The BCP accelerator architecture}
Our solution accelerates BCP execution on FPGA. The architecture processes all clauses in a given partition in parallel and should be able to seamlessly accelerate any DPLL-based algorithm. The BCP engine (Figure \ref{fig:processor_fpga_interface}) consists of a control unit (1), an array of Clause Processors (2) and an implication selector (3).
\par
\begin{figure}
\includegraphics[width=0.5\textwidth]{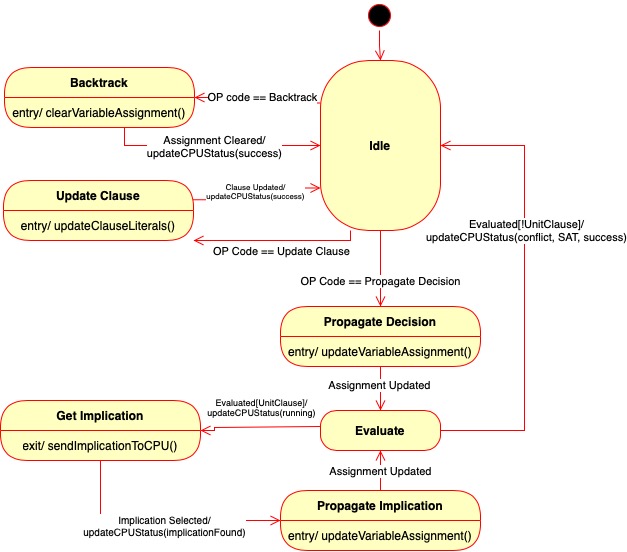}
\centering
\caption{BCP Engine - Control Unit State Diagram.}
\label{fig:bcp_sm}
\end{figure}

 The control unit orchestrates the execution flow of the BCP engine, loading clauses into corresponding processors, clearing backtracking assignments, and broadcasting decisions and implications to clause processors. Behavior is depicted in Figure \ref{fig:bcp_sm}. Clause Processors store clauses as an array of literals; they also contain a copy of the respective variable assignments, updated when a decision or implication is received. Clauses mapped to clause processors can be hot-swapped during runtime, allowing instances to be changed without needing to rebuild the FPGA design. The implication selector is a multiplexer that selects a single implication when multiple clause processors generate implications. Unlike Davis et al \cite{4555925}, our design does not employ an implication conflict detector; instead, the BCP engine propagates the chosen implication and identifies conflicts during the evaluation stage.

\subsection{Formulae partitioning}
Instances of SAT problems may be comprised of an arbitrary number of variables and clauses. Available hardware fabric limits the size of problems solvable on FPGA; large problems must be partitioned. BCP is performed separately on each partition, and implications are propagated to all other partitions. The problem partition currently assigned to the FPGA fabric is controlled by software.
\par The partitioning approach utilized for the results in this paper is a simple greedy approach: a partition is as big as possible, limited by the number of clauses and variables that fit on hardware. Clauses are chosen sequentially, as they appear in the formula to evaluate. Results (see Section \ref{sec:results}) suggest partitioning is the bottleneck in our approach: some partition assignments result in performance improvement, whilst others result in performance degradation, depending on the amount of required partition swapping. Thus, a more selective partitioning algorithm is a logical next step to improve this area of research, but outside the scope of this paper, and reserved for future work.


    

\subsection{Processor-FPGA interface}
The processor communicates with the FPGA through AXI \cite{restuccia2020axi}, sending commands and data by directly writing to hardware registers. Once commands are issued, the processor polls for status updates and new implications.  Status updates informs the processor of conflicts, availability of implications and completion, and dictates the flow of DPLL. The processor stores a copy of implications to avoid re-assigning implied variables, and to propagate to other partitions.

\section{Experiments and Results}\label{sec:results}

We evaluate our accelerator on a Zynq chip, with 14400 LUTs and 28800 FF. Implementation is clocked at 106.66 MHz, utilizing 13151 LUTs, 11059 FFs, and 647 LUTRAM of on-chip memory.

\begin{table}[]
\begin{tabular}{l|lll|}
\cline{2-4}
                                            & \multicolumn{3}{c|}{Millions of BCP/s}                                                                                     \\ \hline
\multicolumn{1}{|l|}{\textbf{SAT Instance}} & \multicolumn{1}{l|}{\textbf{Davis et al    \cite{4555925}}} & \multicolumn{1}{l|}{\textbf{\makecell{Thong et al \\  \cite{6691124}}}} & \textbf{Our Design} \\ \hline
\multicolumn{1}{|l|}{bmc-galileo-8}         & \multicolumn{1}{l|}{40}                    & \multicolumn{1}{l|}{102}                                & 175              \\ \hline
\multicolumn{1}{|l|}{bmc-ibm-12}            & \multicolumn{1}{l|}{33}                    & \multicolumn{1}{l|}{150}                                & 169               \\ \hline
\end{tabular}
\caption{BCP engine throughput (BCPs/s) comparison with related work. Results reflect only full availability of data to BCP engines; i.e., maximum theoretical throughput.}
\label{tab:SOTA}
\end{table}

\begin{figure}
\includegraphics[width=0.48\textwidth]{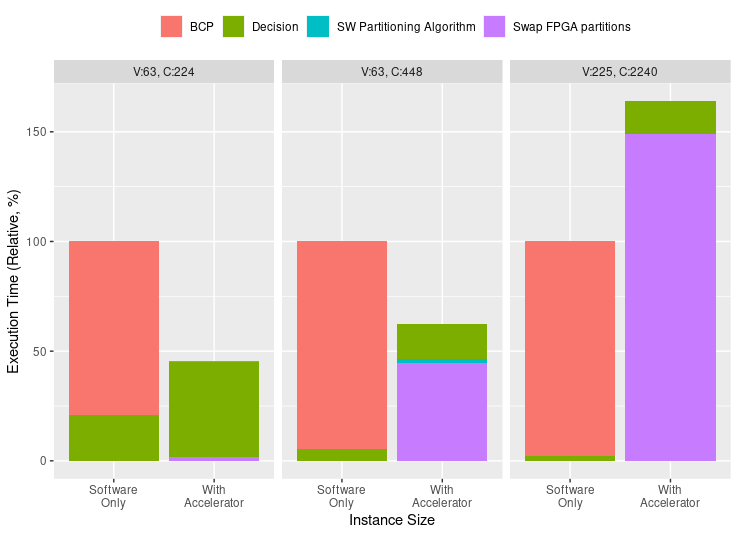}
\centering
\caption{Total execution time breakdown across constituent parts.}
\label{fig:BCPvsVariableNumber}
\end{figure}

\subsection{Comparison to related work}

Related literature only reports throughput on BCP calculation, in millions of BCPs per second, measured solely during periods when data is available to the BCP engine. This metric is not useful, as it excludes software execution and data transfer times which account for a significant portion of total execution time. Nonetheless, equivalent comparison between our design and related work is depicted in Table \ref{tab:SOTA}. A detailed breakdown of total execution time into its constituent parts is depicted in Figure \ref{fig:BCPvsVariableNumber}, contextualizing results of Table \ref{tab:SOTA}.

\subsection{Software acceleration}

To evaluate acceleration gains, we evaluate combinations of clause and variable sizes, and measure total execution time across software-only and hardware-accelerated versions. Relative speedup is depicted in Table \ref{tab:matrix} for meaningful combinations. For each combination, we also depict real throughput, in the form of BCPs/s averaged over total execution time. Notice that entry with 63 variables and 224 clauses represents theoretical upper-limit, where the entire formula fits on the accelerator without partitioning.

\begin{table}[]
\setlength\tabcolsep{1.5pt}
\begin{tabular}{cl|llll|}
\cline{3-6}
\multicolumn{1}{l}{}  &   & \multicolumn{4}{c|}{\textbf{Variables}}\\
\cline{3-6} 
\multicolumn{1}{l}{} &  \textbf{}   & \multicolumn{1}{l|}{\textbf{63}} & \multicolumn{1}{l|}{\textbf{126}} & \multicolumn{1}{l|}{\textbf{252}}                     & \textbf{630} \\ 
\hline
\multicolumn{1}{|c|}{\multirow{4}{*}{\rotatebox[origin=c]{90}{\textbf{Clauses}}}}  & \textbf{224}                                                           & \multicolumn{1}{l|}{\makecell{362M BCP/s\\ 2.2x}}     & \multicolumn{1}{l|}{\makecell{17K BCP/s\\0.17x}}   & \multicolumn{1}{l|}{NA}                         & NA\\ \cline{2-6} 
\multicolumn{1}{|c|}{}                                            & \textbf{448}   & \multicolumn{1}{l|}{\makecell{702K BCP/s\\1.6x}}                    & \multicolumn{1}{l|}{\makecell{21K BCP/s\\0.21x}}   & \multicolumn{1}{l|}{\makecell{13K BCP/s\\0.08x}}   & NA\\ \cline{2-6} 
\multicolumn{1}{|c|}{}                                            & \textbf{2240}  & \multicolumn{1}{l|}{\makecell{441K BCP/s\\1.91x}}                   & \multicolumn{1}{l|}{\makecell{22K BCP/s\\1.26x}}   & \multicolumn{1}{l|}{\makecell{16K BCP/s\\0.61x}}   & \makecell{12K BCP/s\\0.10x} \\ \cline{2-6} 
\multicolumn{1}{|c|}{}                                            & \textbf{22400} & \multicolumn{1}{l|}{\makecell{313K BCP/s\\6.32x}}                   & \multicolumn{1}{l|}{\makecell{20K BCP/s\\5.04x}}   & \multicolumn{1}{l|}{\makecell{16K BCP/s\\4.86x}}   & \makecell{14K BCP/s\\3.31x}  \\ \hline
\end{tabular}
\caption{Relative hardware/software speedup and effective BCP engine throughput for varying clauses/variables sizes.}
\label{tab:matrix}
\end{table}

\par
\begin{figure}[t!]
\includegraphics[width=0.48\textwidth]{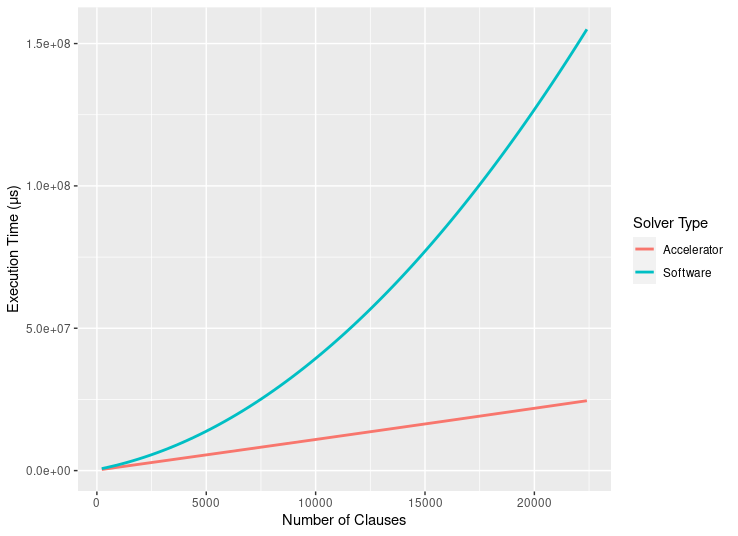}
\centering
\caption{Effect of clauses size increase on total execution time, for 63 variables.}
\label{fig:BCPvsNumOfClauses}
\end{figure}

 To evaluate the different effects of clause/variable sizes on execution, we fix of them and vary only the other, measuring total execution time respectively: results are depicted in Figures \ref{fig:BCPvsNumOfClauses} and \ref{fig:BCPvsVariableNumber}.

\begin{figure}[t!]
\includegraphics[width=0.48\textwidth]{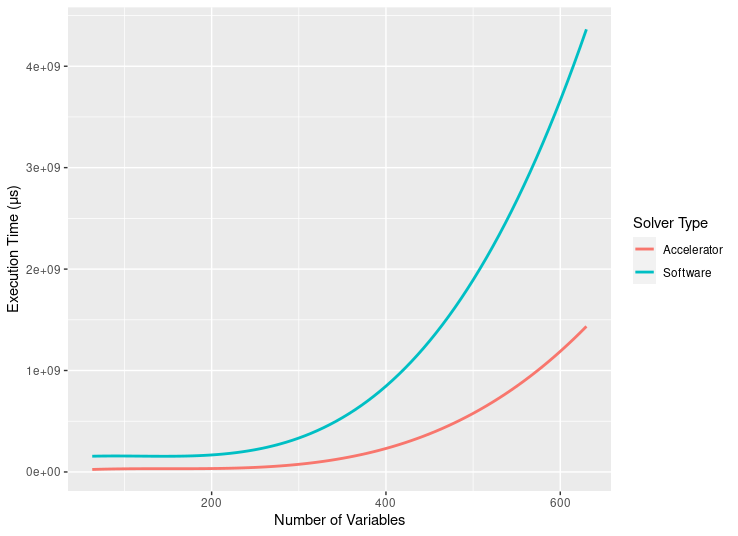}
\centering
\caption{Effect of variables size increase on total execution time, for 22400 clauses.}
\label{fig:BCPvsVariableNumber}
\end{figure}

\section{Conclusion}

A SAT solver accelerator suitable for FPGA-based embedded applications has been described and evaluated. Our approach built on prior work by Davis et al, exploiting fine-grained parallelization. Unlike Davis et al, our solution is not constrained by variable sharing across clauses; instead, our partitioning approach, hot-swapping FPGA assignments at runtime as required for computation, ensures formulas are eventually deemed satisfiable or not. Unlike prior work by Thong at al, which also attempted to eliminate Davis et al's bottleneck, we still take full advantage of FPGA parallelization opportunities: this results in higher theoretical maximum performance (Table \ref{tab:SOTA}) and higher practical performance for a range of clause-variable assignments (Table \ref{tab:matrix}) when FPGA swapping time, as a function of partitioning algorithm, does not exceed benefits of BCP acceleration.
\par Our solution's performance is constrained by the partitioning approach: in all likelihood, minimizing the number of partitions each variable appears in will result in the least amount of swapping time. This, of course, must be balanced with the total number of partitions. An algorithm that partitions a formula in such a way as to achieve pareto-optimality among number of partitions and variable dispersion is not known at this time: future work will focus on this area of research, potentially through stochastic solutions (as exact solutions may well be of the same level of time complexity as SAT).
\par Our artifacts are available in open-source form, and should provide valuable help to practitioners that wish to deploy them in their own solutions.

\section*{Acknowledgments}
We acknowledge the support of the Natural Sciences and Engineering Research Council of Canada (NSERC)

\bibliographystyle{IEEEtran}
\bibliography{IEEEabrv,refs}

\vfill

\end{document}